\documentclass{Interspeech2024}

\usepackage{hyperref}
\usepackage{url}
\usepackage{wrapfig}
\usepackage{booktabs}
\usepackage{tabularx}
\usepackage{multirow}
\usepackage{enumitem}
\usepackage{pifont}
\usepackage{xspace}
\usepackage{subcaption}
\usepackage{graphicx} 
\usepackage{xcolor}
\usepackage{hyperref}

\newcolumntype{L}{>{\raggedright\arraybackslash}X}
\newcommand{\ours}{SESD\xspace}
\newcommand{\ourslong}{Sample-Efficient Speech Diffusion\xspace}

\newcommand\mypara[1]{\vspace{1.5mm}\noindent\textbf{#1}}

\interspeechcameraready

\title{Sample-Efficient Diffusion for Text-To-Speech Synthesis}

\name[affiliation={1,2*}]{Justin}{Lovelace}
\name[affiliation={2}]{Soham}{Ray}
\name[affiliation={2}]{Kwangyoun}{Kim}
\name[affiliation={1,2}]{Kilian Q.}{Weinberger}
\name[affiliation={2,3**}]{Felix}{Wu}

\address{
  $^1$Cornell University, USA\\
  $^2$ASAPP Inc., USA \\
  $^3$Character AI, USA}
\email{jl3353@cornell.edu, \{sray, kkim\}@asapp.com, kqw4@cornell.edu, felix@character.ai}

\begin{document}

\maketitle

\def\thefootnote{*}\footnotetext{Work done during an internship at ASAPP.}\def\thefootnote{\arabic{footnote}}
\def\thefootnote{**}\footnotetext{Work done at ASAPP.}\def\thefootnote{\arabic{footnote}}
\begin{abstract}
    
    This work introduces \textit{\ourslong} (\ours), an algorithm for effective speech synthesis in modest data regimes through latent diffusion. 
    It is based on a novel diffusion architecture, that we call \textit{U-Audio Transformer (U-AT)}, that efficiently scales to long sequences and operates in the latent space of a pre-trained audio autoencoder. Conditioned on character-aware language model representations, \ours  achieves impressive results despite training on less than 1k hours of speech – far less than current state-of-the-art systems. In fact, it synthesizes more intelligible speech than the state-of-the-art auto-regressive model, VALL-E, while using less than 2\% the training data. Our implementation is available at \url{https://github.com/justinlovelace/SESD}.
\end{abstract}

\section{Introduction}
Neural approaches have revolutionized generative speech modeling, with recent advances driven by auto-regressive and diffusion-based systems~\cite{le2023voicebox,wang2023neural}. These improvements, however, come with a cost. Generative models are data hungry, and state-of-the-art systems have used increasingly large volumes of annotated data. This poses challenges for the application of these methods to low-resource domains and languages. Learning effective generative models with limited data has so far remained an open challenge.

To address this data bottleneck, we develop a latent diffusion model that can exploit abundant
 \textit{unlabeled} speech data and therefore requires only a fraction of the \textit{labeled} data~\cite{rombach2021highresolution}.  We utilize a pre-trained autoencoder to map high-dimensional speech waveforms to compact latent representations. By training a diffusion model to generate samples in the lower-dimensional latent space, we offload modeling of fine-grained data characteristics to the unsupervised autoencoder. This allows the diffusion model to focus on the more tractable latent space, thereby improving data efficiency. 

In speech synthesis, the generated audio must align with the text transcript. This makes diffusion models a proper fit, because they can incorporate complex conditioning information into the generative process. However, with limited training data it is challenging to generalize across diverse transcripts. To address this issue, we condition our model on representations from a pre-trained language model. These representations, learned through self-supervised pre-training, contain the rich linguistic information necessary for natural speech synthesis and help our model generalize effectively to diverse text inputs.

Building on these insights, we introduce \textit{\ours} (\ourslong), a sample-efficient latent diffusion framework that achieves impressive results with less than 1k hours of speech data. We develop a diffusion architecture, the U-Audio Transformer (U-AT), that scales efficiently to long audio sequences. It consists of a 1D U-Net that downsamples the audio features before applying a transformer backbone to model global speech characteristics. Crucially, we propose a position-aware cross-attention mechanism to condition the model on representations from a frozen character-aware language model, ByT5-base \cite{xue2022byt5}. To increase our model's alignment with the transcript, we adjust the diffusion loss weighting to emphasize performance at high noise levels where the global structure of the speech (e.g. word placement) is being determined.

With these innovations, \ours can synthesize highly intelligible speech directly from text transcripts, without the explicit phoneme alignment required by current TTS diffusion models \cite{le2023voicebox, shen2023naturalspeech}. For text-only TTS, our framework achieves a word error rate (WER) of 2.3\%, nearly matching the 2.2\% WER of natural human speech. For speaker-prompted synthesis, \ours generates audio with a WER rate of 2.3\% and a speaker similarity score of 0.617, outperforming the state-of-the-art autoregressive model VALL-E (WER 5.9\%, similarity 0.580) which uses 62.5x times more training data \cite{wang2023neural}.

\section{Related Work}
Most related are the diffusion TTS models, NaturalSpeech2 (NS2) \cite{shen2023naturalspeech} and VoiceBox \cite{le2023voicebox}. 
They depend on phonemizers and aligners for frame level phonetic transcripts, which can introduce errors \cite{mcauliffe17_interspeech}. 
Both need phoneme duration annotations for generation, necessitating an external model for phoneme duration prediction. Our system, however, can synthesize varied speech with just the utterance duration and transcript.
NS2 also requires pitch annotations and a speech prompt, unlike our system which supports text-only generation. Importantly, our method is more data-efficient, requiring far less annotated data than NS2 and VoiceBox by 45.8x and 62.6x, respectively.

\section{Background}
Diffusion models \cite{sohl2015deep, ddpm, kingma2021variational} are latent variable models with latents $\mathbf{z}  = \{\mathbf{z}_t | t\in [0,1] \}$ given by a forward diffusion process $q(\mathbf{z}|\mathbf{x})$, which defines a gradual transition from the data distribution, $\mathbf{x} \sim p(\mathbf{x})$, to a Gaussian distribution. The Markovian forward process iteratively adds Gaussian noise to the data over time and satisfies
\begin{align*}
    q(\mathbf{z}_t|\mathbf{z}_s)=\mathcal{N}(\mathbf{z}_t; \alpha_{t|s}\mathbf{z}_s, (1-\alpha_{t|s}^2)\mathbf{I}),\\
    q(\mathbf{z}_t|\mathbf{x}) = \mathcal{N}(\mathbf{z}_t; \alpha_t\mathbf{x}, (1-\alpha_t^2)\mathbf{I})
\end{align*}
where $\alpha_{t|s} = \alpha_t/\alpha_s$
and $0 \leq s < t \leq 1$. The noise schedule, determined by $\alpha_t\in [0,1]$, monotonically decreases the signal-to-noise ratio (SNR), $\lambda_t =\frac{\alpha_t^2}{1-\alpha_t^2}$ as a function of the time, $t$, such that the final latent becomes approximately Gaussian, $q(\mathbf{z}_1) \approx \mathcal{N}(\mathbf{0}, \mathbf{I})$. The forward process therefore defines a transition from the data distribution to a Gaussian distribution.

Diffusion models define a generative process to invert the forward process. This specifies a transition from Gaussian noise, which can be sampled analytically, to the unknown data distribution. Inverting this process can be reduced to learning a \textit{denoising network}, $\hat{\mathbf{x}}_\theta(\mathbf{z}_t, t, \mathbf{c}) \approx \mathbf{x}$, that reconstructs the clean data given some noisy latent, the time, and (optionally) some conditioning information, $\mathbf{c}$, about the data. The conditioning information could be a textual description of an image \cite{saharia2022photorealistic} or, in our case, a textual transcription of some speech. 

In practice, the denoising network is often parameterized as a noise prediction network \cite{ddpm} or a velocity prediction network \cite{salimans2022progressive}, where the velocity is given as $\mathbf{v} = {\alpha_t}\bm{\epsilon} - \sqrt{1-\alpha^2_t} \mathbf{x}$, to improve training stability and performance \cite{salimans2022progressive}.  We adopt the $\mathbf{v}$-parameterization throughout this work and therefore train the denoising network with the regression objective
\[ \mathcal{L}(\theta) = \mathbb{E}_{t,\mathbf{x}, \epsilon} [  w(\lambda_t) \lVert\hat{\mathbf{v}}_{\theta}(\mathbf{z}_t, t, \mathbf{c}) - \mathbf{v} \rVert_2^2 ] \]
with some time-dependent weighting, $w(\lambda_t)$, that is set empirically to emphasize noise levels that are important for downstream perceptual quality \cite{ddpm, nichol2021improved}. This loss function is the weighted variational lower bound of the log likelihood of the data under the forward diffusion process \cite{sohl2015deep, ddpm, kingma2021variational}.

\section{\ourslong}
\begin{figure}[h]
\centering
\includegraphics[width=\linewidth]{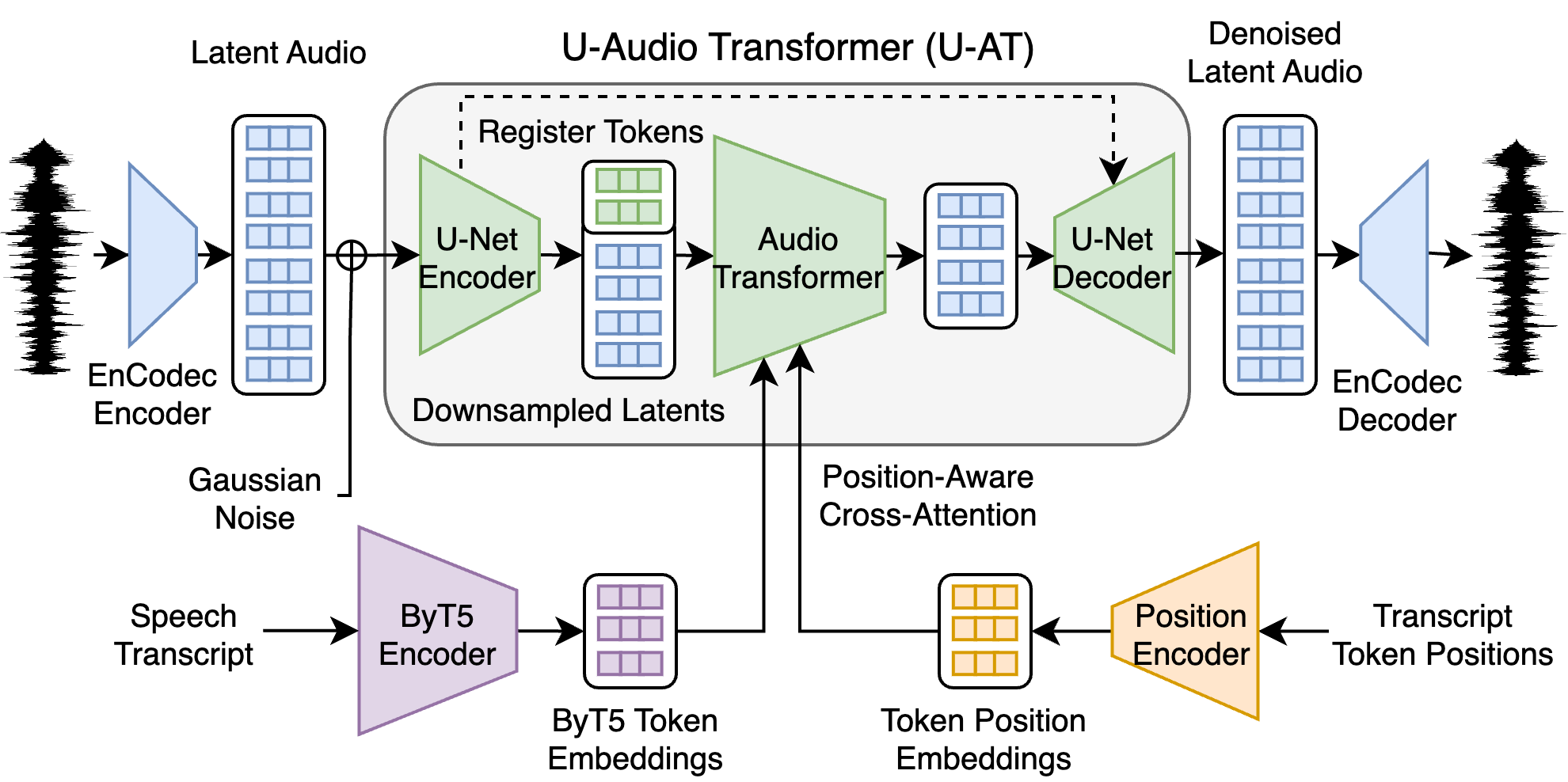}
\caption{\textbf{Overview of our \ourslong architecture}.  \label{fig:u-audt}}
\vspace{-2ex}
\end{figure}

\mypara{Latent Audio Diffusion.}
While auto-regressive approaches require discrete tokens, diffusion models are effective at generating continuous representations. This avoids potential information loss from quantization and and enables modeling long sequences more efficiently. To leverage these benefits, we train our diffusion model on the continuous latent embeddings from a pretrained audio auto-encoder.
Specifically, we utilize the publicly available EnCodec autoencoder to map 24kHz waveforms to sequences of 75 latent vector representations per second of audio \cite{defossez2022highfi}. EnCodec applies residual vector quantization to map each continuous latent vector to multiple discrete tokens capturing increasingly fine details. Instead of modeling these discrete tokens, we train our diffusion model to generate the 75Hz 128-dimensional continuous embeddings from the EnCodec encoder prior to quantization.

This continuous latent diffusion approach significantly reduces the effective sequence length compared to modeling tokens - a 10 second clip consists of just 750 latent vectors rather than 24,000 tokens after quantization (a 32x reduction). The continuous latents generated during inference can then be quantized and decoded by EnCodec to produce the waveform.

\mypara{U-Audio Transformer (U-AT).}
For our diffusion network, we propose the U-Audio Transformer (U-AT), a hybrid architecture that combines the strengths of U-Nets and transformers (see \autoref{fig:u-audt}). U-Nets are well-suited for high-resolution data, while transformers excel at capturing long-range dependencies and incorporating conditioning information.
In the U-AT, we first use a 1D U-Net to downsample the lengthy audio features from a maximum length of 1504 frames to 188 frames. This downsampling step allows us to apply a deep transformer backbone to the compressed sequence, incorporating information from the transcript \cite{vaswani2017attention}. Processing the full-resolution input with a transformer would be computationally prohibitive.

To enhance the transformer's capacity for modeling global information, we incorporate a recent advance from vision transformers \cite{darcet2023vision} and prepend 8 learnable register tokens to the downsampled features. These tokens act as global memory slots, enabling the transformer to better process global information. After applying the transformer, the register tokens are discarded, and the U-Net decoder upsamples only the corresponding audio features back to the original sequence length for the final prediction.
Hybrid U-Net/transformer architectures have shown promise for high-resolution image diffusion \cite{hoogeboom2023simple}, motivating our adaptation to the audio domain.

\mypara{Position-Aware Cross-Attention.}
Properly aligning the generated speech with the input transcript is a critical challenge in text-to-speech synthesis. To improve alignment, we introduce position-aware cross-attention layers in the transformer model that attend to transcript representations from a frozen ByT5-base encoder \cite{xue2022byt5}. To explicitly incorporate positional information about the tokens in the transcript, we introduce a neural Position Encoder that maps the relative positions of the transcript tokens to key vectors. We sum these positional key vectors with the corresponding key vectors from the ByT5 embedding in the cross-attention mechanism. This allows the model to directly search for and attend to the relevant positions within the transcript when generating each audio frame.

Specifically, we compute the cross-attention logits as:
\[\mathbf{A}_{ij}=\mathbf{q}_i^\top (\mathbf{k}_j + f_{\theta}(j/m))\]
where $\mathbf{q}_i \in \mathbb{R}^d$ is the query vector for audio frame $i$, $\mathbf{k}_j \in \mathbb{R}^d$ is the key vector for the ByT5 text embedding $j$ in the sequence of $m$ bytes, and $f_{\theta}(j/m)\in \mathbb{R}^d$ is a relative position embedding computed with a lightweight MLP. This positional encoding is critical for generating speech aligned with the transcript.

\mypara{Diffusion Loss Weighting.}
Properly emphasizing the diffusion noise levels that are most important for perceptual quality is critical~\cite{kingma2023understanding}. Previous work has utilized symmetric weightings like the V-Weighting or unimodal distributions centered around moderate noise levels \cite{kingma2023understanding, le2023voicebox}. However, in text-to-speech synthesis, the input transcript and speech prompt provide valuable signal even at high noise levels where the signal in the corrupted latent itself is limited. These high noise levels are precisely where the conditioning information is most beneficial for resolving the global speech structure and aligning it with the provided transcript.

\begin{figure}[h]
\centering
\includegraphics[width=\linewidth]{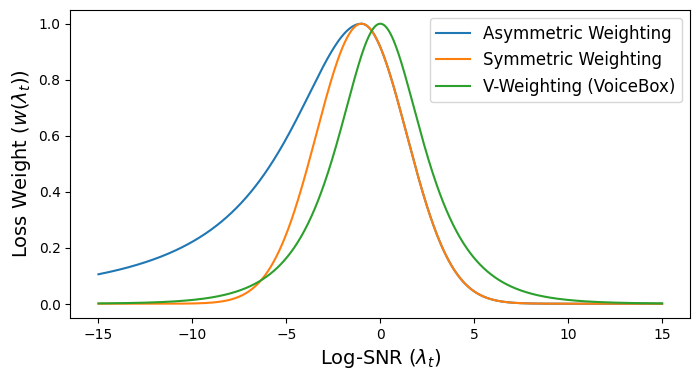}
\caption{\textbf{Diffusion loss weighting across noise levels.} We allocate significant weight to higher levels of noise to improve transcript alignment.
\label{fig:noise_schedules}}
\vspace{-5ex}
\end{figure}

We therefore propose an asymmetric diffusion loss weighting that emphasizes performance at high noise levels where the transcript and prompt are relied upon to estimate the original speech. We visualize our proposed asymmetric weighting, a symmetric weighting baseline, and the V-weighting in \autoref{fig:noise_schedules}. Our proposed weighting dedicates more model capacity to resolving aspects like word placement and positioning compared to symmetric weightings.
Specifically, we parameterize the weighting $w(\lambda_t)$ with a heavy-tailed Cauchy distribution for high noise levels $\lambda_t < -1$, combined with a unimodal normal for lower noise levels:
\begin{align*}
w(\lambda_t)= \begin{cases}
\frac{1}{Z_c} \text{Cauchy}(\lambda_t;-1, 4.8) & \text{if } \lambda_t < -1\\
\frac{1}{Z_n} \mathcal{N}(\lambda_t;-1,2.4) & \text{if } \lambda_t \geq -1
\end{cases}
\end{align*}
where $Z_c$ and $Z_n$ normalize the densities to 1 at $\mu=-1$. This asymmetric weighting improves transcript alignment in our generations compared to symmetric alternatives. For training efficiency, we utilize the adaptive noise scheduler \cite{kingma2023understanding} to reduce loss estimate variance. 

\mypara{Duration Prediction.}
Our approach avoids the need for predicting phoneme durations as an intermediate step, which can introduce errors. During training, we provide the diffusion network with noisy latents of the correct sequence length corresponding to the full utterance duration. At inference time, we only specify the overall duration, not individual phoneme durations. In contrast, models like NaturalSpeech2 and VoiceBox require an external phoneme duration prediction model.

Our model instead learns to resolve the phoneme durations in an end-to-end manner from only the text transcript during diffusion training. For duration prediction at inference time, we simply fine-tune the ByT5-base model as a stochastic duration predictor in a sequence-to-sequence manner. Conditioned on the transcript, it generates utterance durations (e.g. "4.51") auto-regressively with nucleus sampling (p=0.95) \cite{holtzman2019curious}, achieving a 1.4 second RMSE. Importantly, our approach is agnostic to the duration selection method, avoiding cascaded errors from explicit phoneme duration modeling.

\mypara{Speaker-Prompted Generation.}
The ability to perform speaker-prompted generation, where a short sample of reference audio conditions the generation on the desired speaker's voice characteristics, is a valuable capability for TTS systems. Diffusion models can perform speaker-prompted TTS through audio inpainting \cite{le2023voicebox}. We train our denoising network for both text-only and speaker-prompted TTS synthesis in a multi-task fashion. With probability $p=0.5$, we train the network to perform audio inpainting by concatenating a clean audio latent with a noisy latent vector. We sample a duration $d$ and concatenate the start of the latent audio representation $\mathbf{x}[{:}d]$ with the end of the noisy latent $\mathbf{z}_t[d{:}]$ to construct the input. We also introduce a binary embedding to identify corrupted frames, which we sum with the input after the initial projection. When calculating the loss, we mask out frames corresponding to the clean audio. 

For the prompt duration, we sample the proportion of the input, $d \in [0,1] $, to hold out as the clean prompt. For instance, if we sample $d=0.1$ for a 10 second clip of audio, then we use the frames corresponding to the first second of audio as the clean prompt. For sampling the duration, we use a Beta distribution with a mode of $.01$ and a concentration of $5$ to emphasize challenging cases with very short prompts. During inference, we prepend a speaker's reference audio and the associated text to perform speaker-prompted TTS synthesis.

\mypara{Classifier-Free Guidance.}
To enable classifier-free guidance \cite{ho2021classifier}, we drop the text with probability $p=0.1$ and jointly train a conditional and unconditional diffusion model. During inference, we introduce a sampling parameter $w$, and compute 
\[\hat{\mathbf{v}}_{\theta}^w(\mathbf{z}_t, t, \mathbf{c}) = \hat{\mathbf{v}}_{\theta}(\mathbf{z}_t, t) + w*(\hat{\mathbf{v}}_{\theta}(\mathbf{z}_t, t, \mathbf{c}) -\hat{\mathbf{v}}_{\theta}(\mathbf{z}_t, t)).\]
When $w=1.0$, this reduces to the conditional diffusion model, and setting $w>1.0$ increases the influence of the conditioning information.
For the cross-attention layers, we concatenate a learnable null embedding with the text features along the sequence dimension. We mask out the text features to drop conditioning information.

\begin{figure*}[h]
\centering
\includegraphics[width=.95\linewidth]{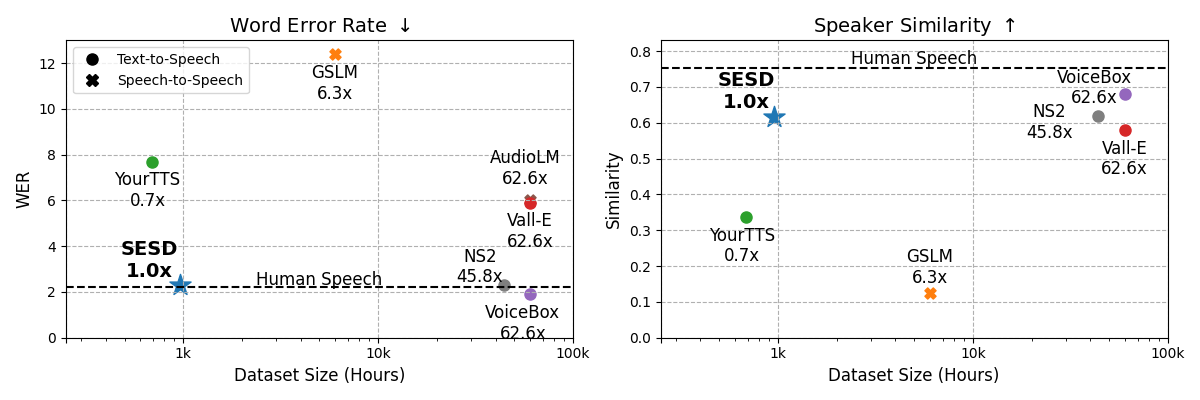}
\caption{\textbf{Speaker-prompted performance across dataset sizes.} We display the relative size of the training dataset for each method.
\label{fig:ablation}}
\vspace{-3ex}
\end{figure*}

\mypara{Implementation Details.}
We begin with the 2D U-Net design used by iDDPM \cite{nichol2021improved} for image diffusion and replace its 2D convolutions with corresponding 1D convolutions to adapt the U-Net to 1D sequences. For instance, we substitute each 2D convolution of size 3x3 with a 1D convolution of size 3. We make similar substitutions for the downsampling and upsampling operations. Our U-Net has 4 stages which downsample the input from 1504 frames to 188 frames. We utilize a feature dimensionality of 512 throughout the network. 

We use a transformer backbone \cite{vaswani2017attention, Peebles2022DiT} with 8 layers and a dimension of 512. We encode positional information with a 1D Dynamic Position Bias (DPB) \cite{wang2021crossformer}. 
This introduces a lightweight MLP that maps relative offsets between locations, $\Delta x_{i,j} \in \{..., -1, 0, 1, 2,...\} $, to head-specific bias terms that are added to the self-attention logits before the softmax.
To condition the diffusion network on the level of noise, we utilize $\alpha$-conditioning \cite{chen2021wavegrad,chen2023importance}. We map $\alpha_t$ to a sinusoidal position embedding \cite{vaswani2017attention} and pass it through an MLP to obtain a time embedding. We condition the U-Net residual blocks and transformer feedforward layers on the time embedding following standard practice from prior image diffusion work \cite{Peebles2022DiT}. 
We pad the audio with silence up to 20 seconds (i.e., 1504 latents), and mask out the silence from the network.

Our final model has 137M trainable parameters. We train \ours for 250k steps with a batch size of 64 utterances on one Nvidia A6000 GPU.  We use the AdamW optimizer \cite{loshchilov2018decoupled} with a 1000-step linear warmup, a peak learning rate of 2e-4, a cosine learning rate decay, and independent weight decay of 2e-4. We apply dropout of 0.1 to the feedfoward, self-attention, and cross-attention layers in the transformer. We compute an exponential moving average (momentum of 0.9999) of the training model. For our ablation study, we train all models for 100k steps without dropout or weight decay. 
For generation, we use 250 sampling steps with the scaled cosine noise schedule with a scale factor of 0.5 \cite{chen2023importance,hoogeboom2023simple}. We use the DDPM sampler \cite{ddpm} with $w=5.0$ for text-only synthesis and the DDIM sampler \cite{ddim} with $w=8.0$ for speaker-prompted synthesis. Full implementation details and model checkpoints are provided in our code release to enable exact reproduction of our work.


\section{Experiments}
We utilize the clean and other training splits of the LibriSpeech (LS) dataset \cite{panayotov2015librispeech}, totaling 960 hours of speech, to train \ours. For evaluation, we follow prior work \cite{wang2023neural, le2023voicebox} and consider a filtered subset of LS test-clean consisting of clips between four and ten seconds in length. For speaker-prompted TTS, we utilize a 3 second clip from another sample of the same speaker.
 
\mypara{Baselines.} For text-only synthesis, we compare against VITS \cite{kim2021conditional}, a variational autoencoder with adversarial training. We consider both VITS variants: the single-speaker VITS-LJ trained on LJ Speech, and the multi-speaker VITS-VCTK trained on VCTK. 
We also compare against English MMS-TTS \cite{pratap2023scaling}, a recent single-speaker model. 
For speaker-prompted TTS, we compare against YourTTS \cite{casanova2022yourtts}, a VITS model conditioned on a speech prompt.
We follow the evaluation protocol of the recent state-of-the-art generative models such as VALL-E \cite{wang2023neural} and VoiceBox \cite{le2023voicebox} and compare against their reported metrics. We also compare against the speech-to-speech baselines GSLM and AudioLM \cite{lakhotia-etal-2021-generative, borsos2023audiolm}.

\begin{wraptable}{r}{0.4\columnwidth}
\centering
\caption{Text-Only TTS \label{tab:text-only}}
\footnotesize
\resizebox{1\linewidth}{!}{
\begin{tabular}{@{}lc@{}}
\toprule
Method & WER $\downarrow$ \\
\midrule
VITS-VCTK \cite{kim2021conditional}& 9.1 \\
VITS-LJ \cite{kim2021conditional} & 4.2 \\
MMS-TTS \cite{pratap2023scaling} & 7.2 \\
\ours (Ours) & 2.3 \\
\midrule
Human Reference & 2.2 \\
\bottomrule
\end{tabular}
}
\end{wraptable}
\mypara{Evaluation Metrics.} To evaluate the \textit{intelligibility} of the synthesized audio, we transcribe the speech with a pre-trained ASR model and compute the WER between the transcribed text and original transcript. We use the HuBERT-L model \cite{hsu2021hubert} employed by prior work \cite{wang2023neural, le2023voicebox}\footnote{\url{https://huggingface.co/facebook/hubert-large-ls960-ft}}. For speaker-prompted TTS, we evaluate the \textit{similarity} between the prompt and synthesized speech by utilizing the pre-trained speaker verification model from prior work\footnote{The WavLM-Large model released at \url{https://github.com/microsoft/UniSpeech/tree/main/downstreams/speaker_verification}.} \cite{wang2023neural, le2023voicebox}. We report the cosine similarity between speaker embeddings for the re-synthesized prompt and synthesized speech.

\section{Results}

Our results in \autoref{tab:text-only} demonstrate that our method can generate intelligible speech in a text-only setting, nearly matching the word error rate of the ground truth audio. 
Our text-only WER surpasses that of the single-speaker models while providing the additional capability of multi-speaker synthesis. In the speaker-prompted setting, our model generates speech that maintains the characteristics of the prompt. Notably, \ours outperforms the SoTA auto-regressive system, VALL-E, in terms of both the WER and the neural speaker similarity metric, with less than 2\% the training data. We also match the performance of the latent diffusion NS2 system using 2.2\% of the training data. 
We demonstrate the importance of our various design decisions in \autoref{fig:ablation}. Our position-aware cross attention mechanism, model architecture, text encoder, and diffusion loss weighting are critical for generating intelligible speech.

\begin{figure}[h]
\centering
\includegraphics[width=\linewidth]{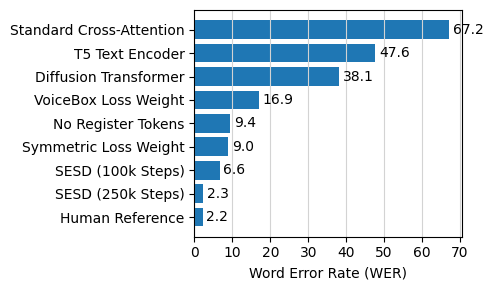}
\caption{\textbf{Ablation studies.} 
\label{fig:ablation}}
\vspace{-4ex}
\end{figure}

\section{Conclusion}
We present \ours, a highly sample-efficient latent diffusion framework for text-to-speech synthesis that achieves strong results in a modest data regime. The key ingredients in the success of \ours are: a novel diffusion architecture that efficiently models long audio sequences, incorporating representations from a byte-level language model that capture linguistic properties critical for natural speech synthesis, and modifying the diffusion loss weighting to improve text-speech alignment.
Together, these innovations enable \ours to perform speech synthesis directly from text without explicit phoneme alignment. \ours generates intelligible speech near human-level word error rates with less than 1k hours of training data.

\bibliographystyle{IEEEtran}
\bibliography{mybib}

\end{document}